\documentclass[nofootinbib,twocolumn,showpacs,preprintnumbers,pre,aps]{revtex4-1}

\usepackage{graphicx}
\usepackage{bm}
\usepackage{amsmath}
\usepackage{amssymb}
\usepackage{color}

\begin{document}

\title{Shear viscosity of two-state enzyme solutions}

\author{Yuto Hosaka}

\author{Shigeyuki Komura}\email{komura@tmu.ac.jp}

\affiliation{
Department of Chemistry, Graduate School of Science,
Tokyo Metropolitan University, Tokyo 192-0397, Japan}

\author{David Andelman}\email{andelman@tauex.tau.ac.il}

\affiliation{
Raymond and Beverly Sackler School of Physics and Astronomy,\\
Tel Aviv University, Ramat Aviv, Tel Aviv 69978, Israel}

\date{December 31, 2019}

\begin{abstract}
We discuss the shear viscosity of a Newtonian solution of catalytic enzymes and substrate molecules.
The enzyme is modeled as a two-state dimer consisting of two spherical domains connected with an elastic spring.
The enzymatic conformational dynamics is induced by the substrate binding and such a process 
is represented by an additional elastic spring.
Employing the Boltzmann distribution weighted by the waiting times of enzymatic species in each catalytic cycle, we obtain 
the shear viscosity of dilute enzyme solutions as a function of substrate concentration and its physical properties.
The substrate affinity distinguishes between fast and slow enzymes, and the corresponding viscosity expressions are obtained.
Furthermore, we connect the obtained viscosity with the diffusion coefficient of a tracer particle in enzyme solutions.
\end{abstract}

\maketitle

\section{Introduction}
\label{sec:introduction}

Molecular enzymes are nanometer-size proteins that catalyze chemical reactions in the presence of \textit{substrate} 
molecules.
Here substrates are chemical species that react with enzymes and generate product molecules.
Catalytic processes that are carried out by molecular enzymes in the cytoplasm and the membrane are essential 
for cellular metabolism and homeostasis~\cite{Albertsbook}.
In the presence of a substrate, enzymes undergo conformational changes in each turnover cycle of the chemical 
reaction~\cite{Gerstein1994}.
In order to mimic actual enzymes, these conformational dynamics have been simulated using elastic network models~\cite{Togashi2007,Sakaue2010,Echeverria2011}, and 
the relationship between conformational dynamics and the chemical reaction stages has been studied recently~\cite{Aviram2018}.

One of the long-standing and interesting questions in the field is whether a single enzyme exhibits a motile behavior~\cite{Zhang2019}.
Thanks to recent developments of experimental techniques, diffusion phenomena in enzyme solutions have 
been studied by several groups.
Using fluorescence correlation spectroscopy, Muddana \textit{et al.}~\cite{Muddana2010} reported that diffusion of a single enzyme is enhanced in presence of a substrate.
Later on, Riedel \textit{et al.}~\cite{Riedel2015} showed that the heat released during turnovers also enhances the enzyme diffusion.
Illien \textit{et al.}~\cite{Illien2017_2} however, revealed experimentally that not only exothermic enzymes but also endothermic ones contribute to the diffusion enhancement.
In the presence of a gradient in substrate concentrations, enzymes exhibit collective motions in the direction 
of higher or lower concentrations~\cite{Sengupta2013,Jee2018}.
Moreover, the enhanced diffusion of passive objects in enzymatic solutions have been observed 
independently~\cite{Zhao2017,Dey2016}.

To understand these experimental findings, several models have been proposed using equilibrium as well as non-equilibrium approaches.
Illien \textit{et al.}~\cite{Illien2017} modeled an enzyme consisting of hydrodynamically coupled subunits, and 
introduced two discrete equilibrium states corresponding to a free enzyme and a substrate-enzyme complex.
They showed that diffusion of an enzyme is enhanced due to equilibrium 
fluctuations~\cite{Illien2017,Adeleke-Larodo2019}.
Within a non-equilibrium framework, Golestanian~\cite{Golestanian2015} proposed four possible mechanisms leading to diffusion enhancement by enzymes.
They included self-thermophoresis, boost in kinetic energy, stochastic swimming, and 
collective heating.
Mikhailov and Kapral~\cite{Mikhailov2015,Kapral2016} modeled an enzyme as an active force dipole that exerts forces on the surrounding fluid.
When such dipoles are immersed in aqueous fluids, hydrodynamic collective effects due to force dipoles can lead to 
diffusion enhancement~\cite{Mikhailov2015,Kapral2016,Hosaka2017}.

In spite of these extensive studies on enzyme diffusion, 
a recent experimental work pointed out the difficulty of accounting quantitatively for the observed 
enhanced diffusion within such models as above~\cite{Xu2019}.
Moreover, recent experiments did not observe any change in the diffusion behavior for a specific enzyme that was previously reported to exhibit enhanced diffusion~\cite{Zhang2018,Guenther2019}.
It was also noticed that the viscosity of enzyme solutions is locally reduced while a specific enzymatic reaction 
is taking place~\cite{Zhang2019,Armoskaite2012}.
However, the effect of enzyme conformational changes on the solution shear viscosity has not been considered theoretically despite its importance.

In this paper, we present an analytical study on the shear viscosity of a dilute enzyme solution under steady shear flow.
As a coarse-grained model of catalytic enzymes, we use the two-state dimer model in which conformational changes are induced by substrate binding and product release~\cite{Mikhailov2015}.
Our two-state dimer model consists of two hard spheres representing enzymatic domains, which are connected by a harmonic spring~\cite{Mikhailov2015,Hosaka2019,Flechsig2019}.
Assuming that the conformational distribution is given by the Boltzmann distribution function, weighted by the waiting time of an enzyme, we obtain analytically the shear viscosity of a two-state dimer solution as a function of the substrate concentration.
As a result of the competition between the energy difference of the enzyme two internal states and the 
substrate concentration, we find that the enzyme solution viscosity exhibits a non-monotonic behavior that depends on the physical properties of the binding substrates.
We shall also connect the obtained viscosity with the diffusion coefficient of a tracer particle in enzyme solutions.

The outline of our manuscript is the following.
In Sec.~\ref{sec:viscosity}, we review the derivation of the shear viscosity of dimer solutions originally used to describe polymer solutions.
In Sec.~\ref{sec:model}, we discuss the shear viscosity of a two-state dimer solution that represents enzyme solutions. 
We first introduce the two-state dimer model and discuss the conformational distribution function of dimers. 
Analytical results for the shear viscosity due to dimers and its limiting expressions are presented.
Finally, some discussions and a summary are given in Sec.~\ref{sec:discussion}.

\section{Viscosity of dimer solutions}
\label{sec:viscosity}

\subsection{Shear viscosity}

\begin{figure}[tbh]
\begin{center}
\includegraphics[scale=0.4]{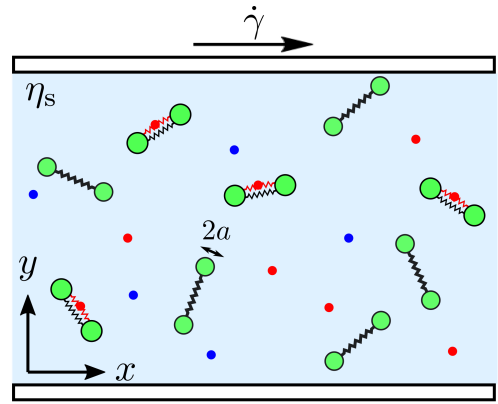}
\end{center}
\caption{
(Color online)
A dilute solution of two-state dimers under steady shear flow with shear rate $\dot{\gamma}$.
Dimers consist of two green spheres of radius $a$ connected with an elastic spring, and immersed in a Newtonian fluid having viscosity 
$\eta_{\rm s}$.
The enzymatic reaction, in which a dimer, a substrate (red circle) and a product (blue circle) participated, is explained 
in Fig.~\ref{fig:cycle}.
}
\label{fig:shear}
\end{figure}

We consider a dilute solution of dimers under steady shear flow as schematically depicted 
in Fig.~\ref{fig:shear}.
Here the solvent viscosity is $\eta_{\rm s}$ and each dimer is composed of two rigid spheres of radius $a$, which are connected by an elastic spring.
The positions of two spheres are denoted by the three-dimensional vectors $\mathbf{r}_1$ and $\mathbf{r}_2$. 
Then, the force acting between the two spheres within the dimer is given by 
\begin{align}
f_\alpha = -\frac{\partial U(r)}{\partial r_\alpha}, 
\end{align}
where $U(r)$ is the elastic potential energy, $r = \vert \mathbf{r} \vert =\vert \mathbf{r}_2-\mathbf{r}_1 \vert$
is the distance between the two spheres, and $r_\alpha$ is the $\alpha$-component of the vector $\mathbf{r}=(r_x,r_y,r_z)$.

In the presence of potential forces, the equation of motion of an overdamped dimer can 
be written as~\cite{Birdbook,Doibook}
\begin{align}
\frac{\partial r_\alpha}{\partial t}=\frac{2}{\zeta}f_\alpha -\frac{2k_{\rm B}T}{\zeta}\frac{\partial \ln\psi}{\partial r_\alpha}
+d_{\alpha\beta}r_\beta,
\label{eq:velocity}
\end{align}
where $\zeta$ is the friction coefficient of the sphere, $k_{\rm B}$ is Boltzmann constant, 
$T$ is the temperature, $\psi(\mathbf{r},t)$ is the time-dependent configurational distribution of a dimer, and
the velocity gradient tensor is given by 
\begin{align}
d_{\alpha\beta}= \frac{\partial v_\alpha}{\partial r_\beta}.
\end{align}
Notice that $v_\alpha$ is the $\alpha$-component of the velocity $\mathbf{v}=(v_x,v_y,v_z)$.
Throughout this work, we assume summation over repeated indices.
The second and third terms on the right-hand side of Eq.~(\ref{eq:velocity}) represent the velocity 
due to thermal motion of the solvent and that imposed by the flow field, respectively.

Such models of dimers have been used extensively to model polymer solutions.
For polymer solutions, the stress tensor due to the presence of dimers is given~\cite{Birdbook,Doibook}
\begin{align}
\sigma_{\alpha\beta} = n\langle r_\alpha f_\beta\rangle,
\label{eq:stress}
\end{align}
where $n$ is the number density (per unit volume) of dimers, and $\langle \cdots\rangle$ denotes the thermal average 
over all dimer configurations.
To calculate the statistical average in Eq.~(\ref{eq:stress}), we introduce the following 
Fokker-Planck equation for the conformational distribution $\psi(\mathbf{r},t)$
\begin{align}
\frac{\partial \psi}{\partial t} = - \frac{\partial}{\partial r_\alpha} \left( \frac{2}{\zeta}f_\alpha\psi
-\frac{2k_{\rm B}T}{\zeta}\frac{\partial\psi}{\partial r_\alpha} + d_{\alpha\beta}r_\beta\psi \right).
\label{eq:FP}
\end{align}
In the above, the continuity equation
\begin{align}
\frac{\partial \psi}{\partial t} = - \boldsymbol{\nabla} \cdot
\left( \frac{\partial \mathbf{r}}{\partial t}\psi \right),
\end{align}
where $\boldsymbol{\nabla}=(\partial r_x,\partial r_y,\partial r_z)$ and Eq.~(\ref{eq:velocity}) have been used.
From the time evolution of $\langle r_\alpha r_\beta \rangle$ in a steady state, the stress tensor in Eq.~(\ref{eq:stress}) can be written as~\cite{Birdbook,Doibook}
\begin{align}
\sigma_{\alpha\beta} = nk_{\rm B}T\delta_{\alpha\beta} + \frac{n\zeta}{4} \Bigl[d_{\alpha\gamma} \langle  r_\beta r_\gamma\rangle
+ d_{\beta\gamma} \langle r_\alpha r_\gamma \rangle \Bigl].
\label{eq:shear}
\end{align}

For simple shear flow whose velocity components are given by 
$v_x=\dot{\gamma}r_y$, $v_y=v_z=0$, where $\dot{\gamma}$ is the shear rate 
(see Fig.~\ref{fig:shear}), the viscosity due to dimers has a simple form
\begin{align}
\eta = \frac{\sigma_{xy}}{\dot{\gamma}} = \frac{n\zeta}{4} \langle r_y^2 \rangle.
\label{eq:eta_simple}
\end{align}
In order to calculate the average $\langle r_y^2 \rangle$, we need to specify the conformational distribution function $\psi(\mathbf{r})$.

\subsection{Fraenkel dimer model}

Let us first discuss a dimer consisting of two spheres that are connected by a harmonic spring having an elastic constant $K_0$, and a natural length $\ell_0$.
Its potential energy is then given by  
\begin{align}
U_0(r) =\frac{K_0}{2} (r -\ell_0)^{2}.
\label{eq:simpleenergy}
\end{align}
This is the ``\textit{Fraenkel dimer model}"~\cite{Fraenkel1952}, and is different than other polymer dynamic models, 
such as the Hookean dimer model.
For Fraenkel dimers, the conformational distribution function, $\psi_0$, is given by
\begin{align}
\psi_{\rm 0}(r) &= C\exp\left[
-\frac{K_0}{2k_{\rm B}T}(r -\ell_{0})^{2} \right],
\label{passive}
\end{align}
where $C$ is the normalization constant.
Here, we assume that the characteristic relaxation time of a dimer is much smaller than that of a shear flow, 
i.e., $\zeta\ell_0^2/(k_{\rm B}T)\dot{\gamma}\ll1$.
The physical meaning of this condition will be separately explained in Sec.~\ref{sec:discussion}.

Although the shear viscosity of the Fraenkel dimer model was discussed in Ref.~\cite{Bird1997}, its explicit expression 
was not derived.
By calculating $\langle r_y^2 \rangle$ in Eq.~(\ref{eq:eta_simple}) using Eq.~(\ref{passive}), we obtain the shear 
viscosity for a Fraenkel dimer solution $\eta_0$ as
\begin{align}
& \frac{\eta_{\rm 0}(\epsilon)}{G\tau}  = \frac{2\epsilon}{3} 
\nonumber \\
& \times  \frac{2\epsilon(5+2\epsilon)e^{-\epsilon} + \sqrt{\pi\epsilon} (3+12\epsilon+4\epsilon^2)
\left[ 1+{\rm erf}(\sqrt{\epsilon}) \right]}
{ 4\epsilon^2 e^{-\epsilon} + 2\sqrt{\pi\epsilon} (\epsilon+2\epsilon^2)
\left[ 1+{\rm erf}(\sqrt{\epsilon}) \right] },
\label{eq:eta_F}
\end{align}
where $\epsilon=K_0\ell_0^2/(2k_{\rm B}T)$ is the dimensionless 
elastic energy, $G=nk_{\rm B}T$ is the relaxation modulus, 
$\tau=\zeta/(4K_0)$ is the relaxation time, and 
${\rm erf}(x)=(2/\sqrt{\pi}) \int_0^x dt\, e^{-t^2}$ is the error function~\cite{Abramowitz1972}.
Notice that $G\tau$ corresponds to the viscosity of a dimer solution when the natural length of the spring vanishes, 
i.e., $\epsilon=0$~\cite{Bird1997,Doibook}.

The limiting behaviors of $\eta_{\rm 0}$ for the Hookean, $\epsilon\ll1$, and stiff Fraenkel dimers, $\epsilon\gg1$, are given by~\cite{Bird1997,Birdbook}
\begin{align}
\frac{\eta_{\rm 0}(\epsilon)}{G\tau} = \left\{ \begin{array}{lr}{\displaystyle 1 + \frac{4}{3}\sqrt{\frac{\epsilon}{\pi}} } & {~~~\epsilon\ll1}, \\[2.0ex]
 {\displaystyle \frac{2}{3}\epsilon } & {~~~\epsilon\gg1}.
\end{array} \right.
\label{eq:eta0}
\end{align} 
For $\epsilon\ll1$, the viscosity is almost constant, indicating that thermal energy dominates 
over elastic energy.
For $\epsilon\gg1$, on the other hand, the viscosity increases linearly with $\epsilon$.

\section{Two-state dimer solutions}
\label{sec:model}

\subsection{Two-state dimer model}

\begin{figure}[tbh]
\begin{center}
\includegraphics[scale=0.3]{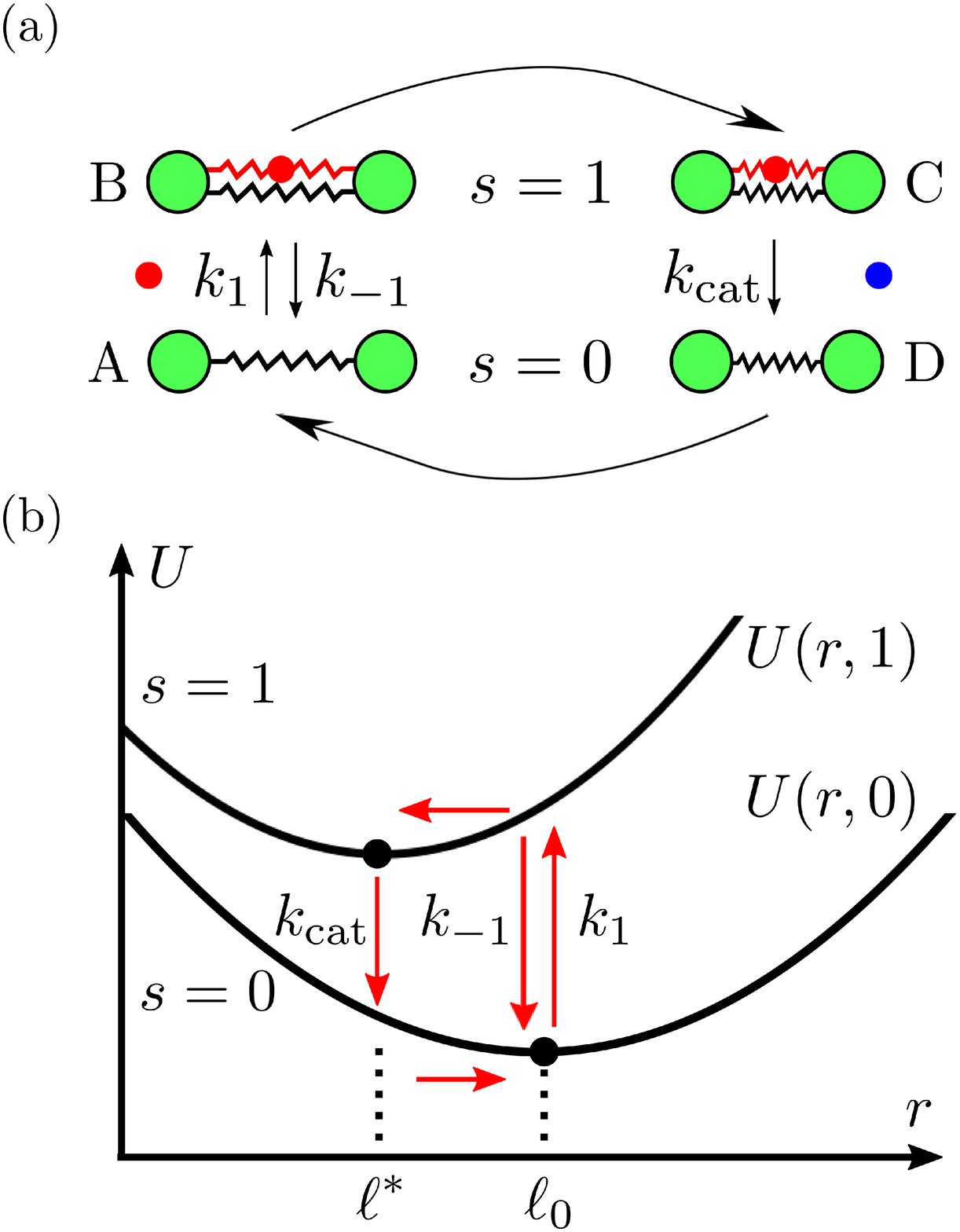}
\end{center}
\caption{
(Color online)
(a)
The enzymatic cycle of two-state dimer model.
A substrate (red circle) binds to a free enzyme ($s=0$) with the reaction rate $k_1$ (A$\rightarrow$B), while its dissociation also occurs with the reaction rate $k_{-1}$ (B$\rightarrow$A).
Once the substrate-enzyme complex ($s=1$) is formed, it starts to contract until the equilibrium conformation is attained (B$\rightarrow$C).
Then, the product (blue circle) is irreversibly released with the reaction rate $k_{\rm cat}$, and the bare enzyme comes back to its initial conformation.
(b)
The schematic illustration of the energy for a two-state dimer as described by Eq.~(\ref{eq:energy}).
There are two energy branches $U(r,0)$ and $U(r,1)$.
The transition between them takes place at $r=\ell_0$ and $r=\ell^\ast$, which are the equilibrium values of 
$U(r,0)$ and $U(r,1)$, respectively, as are indicated by black circles.
This transition is followed by the downhill relaxational motion along each branch. 
The forward and reverse transition rates, $(s = 0)\rightleftarrows(s=1)$, are given by $k_1$,$k_{-1}$, respectively, and $(s = 1)\to (s=0)$ is given by $k_{\rm cat}$.
\label{fig:cycle}
}
\end{figure}

Catalytic enzymes undergo conformational changes in presence of substrate molecules.
To model such situations, we use a previously proposed two-state dimer model with a state parameter that can get two values, 
$s=0$ or $1$~\cite{Mikhailov2015,Flechsig2019,Hosaka2019}.
In Fig~\ref{fig:cycle}(a), we schematically illustrate an enzymatic cycle that is driven by binding a substrate to an enzyme.
In the $s=0$ state, i.e., the state of the dimer with the elastic constant $K_0$ and the natural length of the spring $\ell_0$, 
this model corresponds to the Fraenkel dimer model.

When a substrate is supplied to a dimer enzyme whose size is $r=\ell_0$, a transition from $s=0$ to $s=1$ occurs with the reaction rate $k_1$.
At the same time, the reverse reaction, namely, the substrate dissociation process, can occur also when $r=\ell_0$
with the reaction rate $k_{-1}$.
For the state $s=1$, the substrate adds another intra-dimer interaction, which is modeled as an additional spring, whose elastic constant
and natural length are $K_1$ and $\ell_1$, respectively.
Then, the dimer relaxes to a new equilibrium conformation having the size $r=\ell^\ast$, as will be explicitly given after 
Eq.~(\ref{eq:energy}).
Once the substrate molecule is irreversibly converted to a product molecule with the reaction rate $k_{\rm cat}$, a transition from 
$s=1$ to $s=0$ takes place at $r=\ell^\ast$.
Finally, the product is released from the enzyme.

Notice that the reaction rates, $k_1$, $k_{-1}$ and $k_{\rm cat}$ are the bare rate constants that do not 
depend on the energy difference between any two states. 
This also holds for the reaction rates in the cascade reactions discussed in Appendix~\ref{app:general_psi}.
Moreover, the transition of a dimer occurs only when $r=\ell_0$ or $r=\ell^\ast$; hence, the reaction rates 
$k_1$, $k_{-1}$ and $k_{\rm cat}$ are simply taken to be constant in our model.

The state-dependent total potential energy of this two-state dimer can be written as 
\begin{align}
U(r,s) =\frac{K_0 }{2}(r -\ell_0)^{2} +\frac{sK_1}{2}(r -\ell_1)^2,
\label{eq:energy}
\end{align}
which gives the equilibrium length for $s=1$ as $\ell^\ast=(K_0\ell_0+K_1\ell_1)/(K_0+K_1)$.
In Fig.~\ref{fig:cycle}(b), we schematically illustrate the energy of a two-state dimer given by Eq.~(\ref{eq:energy}) 
when $\ell_0>\ell_1$.
Under this condition, the substrate-enzyme complex shrinks as compared to the bare enzyme~\cite{Mikhailov2015,Flechsig2019,Hosaka2019}.
In this work, however, we do not require such a condition.
In physiological conditions, the sizes of actual substrate-enzyme complexes either decrease ($\ell_0>\ell_1$) or increase ($\ell_1>\ell_0$) upon substrate binding~\cite{Zhang2019}.
Hereafter, the subscripts ``0" and ``1" denote physical values for the enzyme and the substrate-enzyme complex, respectively.

As represented by the second term in the r.h.s.\ of Eq.~(\ref{eq:velocity}), a dimer in our model undergoes conformational 
fluctuations due to thermal energy.
In other words, a free enzyme (or a substrate-enzyme complex) fluctuates around $r=\ell_0$ (or $r=\ell^\ast$) during 
turnover cycles.
This corresponds to the situation in which enzymes are subject to thermal motion of solvent molecules.
Notice, however, that conformational fluctuations between multi-state enzymes~\cite{Kou2005,English2006} are not considered.
This is because the original dimer model~\cite{Mikhailov2015,Flechsig2019,Hosaka2019} that we employ follows 
the simple Michaelis-Menten kinetics [see Eq.~(\ref{eq:MM})] with the advantage that the problem becomes tractable.

\subsection{Conformational distribution function}

The above two-state dimer model describes a chemical equation following the standard Michaelis-Menten reaction~\cite{Michaelis1913}:
\begin {align}
{\rm E} + {\rm S} 
\overset {~k_{\rm 1}~}{\underset {~k_{\rm -1}~}{\rightleftarrows }} {\rm ES }
\xrightarrow {k_{\rm cat}}{\rm E_\ast} + {\rm P}.
\label{eq:MM}
\end{align}
This chemical reaction equation describes the enzymatic cycle composed of three states of an enzyme: a free enzyme (${\rm E}$), a substrate-enzyme complex (${\rm ES}$), and a free enzyme after the 
reaction (${\rm E}_\ast$), as depicted in Fig.~\ref{fig:cycle}.
Furthermore, ${\rm S}$ and ${\rm P}$ stand for the substrate and product, respectively.
When dimers are connected by elastic springs, the time spent during the transition between these chemical states can be characterized by a relaxation time $\tau=\zeta/(4K_0)$ as introduced after Eq.~(\ref{eq:eta_F}).

For a two-state dimer, we assume that the characteristic relaxation time is much smaller than that of a shear flow, i.e., $\zeta\ell_0^2/(k_{\rm B}T)\dot{\gamma}\ll1$ as adopted for the Fraenkel dimer model in Sec.~\ref{sec:viscosity}.
We further assume that the transition time spent between enzymatic states is much smaller than the waiting time in each of the states, $s=0,1$, i.e., $\tau/W_s\ll1$, where the waiting time $W_s$ will be defined later
in Eq.~(\ref{eq:p_i}).
This assumption is justified for enzymes such as adenylate kinase having a relatively large waiting time,
$\tau/W_1\approx0.1$~\cite{Aviram2018}.
For completeness, however, the general case of arbitrary waiting times is discussed in Sec.~\ref{sec:discussion}.
Under these conditions, we can introduce the Boltzmann distribution function that is weighted only by the waiting time in the respective enzymatic states.
The validity of this assumption has been confirmed by numerical solutions of the Langevin equation for a 
single two-state dimer~\cite{Hosaka2019}.

The distribution function for the two-state dimer model for an enzyme is then given by
\begin{align}
\psi_{\rm e}(r) &=
\frac{W_0 e^{-\beta U(r,0)}+ W_1 e^{-\beta U(r,1)}}
{ \int d\mathbf{r}\, \left[ W_0 e^{-\beta U(r,0)} + W_1 e^{-\beta U(r,1)} \right] },
\label{eq:psi}
\end{align}
where $\beta=1/(k_{\rm B}T)$.
Here the waiting time in the state $s$ is defined by~\cite{vanKampenbook,Cao2011}
\begin{align}
W_s&=\int_0^\infty dt\, p_s(t),
\label{eq:p_i}
\end{align}
where $p_s(t)$ is the time-dependent probability distribution function of an enzyme in state $s$, which will be explicitly 
given in Eq.~(\ref{eq:p0p1}).
The case of a cascade reaction containing $N$ substrate-enzyme complexes is discussed in Appendix~\ref{app:general_psi} as a generalization, and Eq.~(\ref{eq:psi}), hence, corresponds to the case $N=1$.

\subsection{Waiting times}

Since we consider a dilute solution of two-state dimers, we employ a single enzyme kinetics to obtain the waiting time that an enzyme spends at each catalytic step (see also Appendix~\ref{app:MM_single}).
The validity of using a single enzyme kinetics for an enzyme solution will be discussed later in this subsection.
For two-state dimers, the corresponding kinetic equations are written in terms of the probability functions as~\cite{Lu1998,Xie2001,Kou2005,English2006}
\begin{align}
\frac{dp_{\rm 0}}{dt} & =  k_{-1} p_{\rm 1} - k_1^\prime p_{\rm 0},
\nonumber \\
\frac{dp_{\rm 1}}{dt} &=  k_{1}^\prime p_{\rm 0} - (k_{-1}+k_{\rm cat}) p_{\rm 1},
\nonumber \\
\frac{dp_{\ast}}{dt}  &=  k_{\rm cat} p_{\rm 1}. 
\label{eq:dp01e0}
\end{align}
Here, $p_0(t)$,  $p_1(t)$ and $p_{\ast}(t)$ are the probability distribution functions for 
the two-state dimer in one of the two states, $s=0$, $1$, and the free enzyme after the catalysis (${\rm E_\ast}$), respectively.
In the above, we have introduced the pseudo first-order rate constant $k_1^\prime=k_1c_{\rm S}$, where
$c_{\rm S}$ is the time-independent substrate concentration.
Such an assumption is justified when $c_{\rm E} \ll c_{\rm S}$ is satisfied, where $c_{\rm E}$ is the enzyme concentration.

By solving the above coupled kinetic equations using the initial conditions, $p_0(0)=1$ and 
$p_1(0)=p_\ast(0)=0$, under the normalization condition  
$p_{\rm 0}(t)+p_{\rm 1}(t)+p_{\ast}(t)=1$, the time-dependent probability distributions 
are obtained~\cite{Lu1998}
\begin{align}
p_{\rm 0}(t) &=  \frac{1}{2a} \left[ \left( a+b-k_1^\prime\right) e^{(a-b)t} 
+\left( a-b+k_1^\prime \right) e^{-(a+b)t}\right],  \nonumber \\
p_{\rm 1}(t)  &= \frac{k_1^\prime}{2a} \left[  e^{(a-b)t} - e^{-(a+b)t} \right], \nonumber \\ 
p_{\ast}(t) &=  \frac{k_1^\prime k_{\rm cat}}{2a} \left[  \frac{1}{a-b}e^{(a-b)t} + 
\frac{1}{a+b}e^{-(a+b)t} \right] + 1,
\label{eq:p0p1}
\end{align}
where 
\begin{align}
a&=\left[(k_1^\prime+k_{-1}+k_{\rm cat})^2/4-k_1^\prime k_{\rm cat}\right]^{1/2}, \nonumber \\
b&=(k_1^\prime+k_{-1}+k_{\rm cat})/2.
\end{align}
Because $a-b<0$ and $a+b>0$, both $p_{\rm 0}(t)$ and $p_{\rm 1}(t)$ decay exponentially for $t\rightarrow\infty$, and consequently $p_\ast\rightarrow1$.

Substituting $p_{\rm 0}(t)$ and $p_{\rm 1}(t)$ of Eq.~(\ref{eq:p0p1}) into Eq.~(\ref{eq:p_i}), we obtain the waiting times 
for $s=0$ and $1$ as 
\begin{align}
W_0 = \frac{k_{-1}+k_{\rm cat}}
{k_1^\prime k_{\rm cat}},~~~~~
W_1 = \frac{1}{k_{\rm cat}}.
\label{P1}
\end{align}
As a result, the distribution function in Eq.~(\ref{eq:psi}) can be written as 
\begin{align}
\psi_{\rm e}(r) &=
 \frac{ e^{-\beta U(r,0)}+ \nu  e^{-\beta U(r,1)}}
{\int d\mathbf{r}\, \left[e^{-\beta U(r,0)} + \nu \, e^{-\beta U(r,1)} \right] },
\label{eq:psi_r}
\end{align}
where we have introduced the dimensionless parameter $\nu$ 
\begin{align}
\nu=\frac{k_1}{k_{-1}+k_{\rm cat}} c_{\rm S} =\frac{c_{\rm S}}{K_{\rm M}}, 
\label{eq:nu}
\end{align}
and $K_{\rm M}$ is the Michaelis constant~\cite{Albertsbook}  
\begin{align}
K_{\rm M}=\frac{k_{-1}+k_{\rm cat}}{k_1}.
\label{eq:MMconst}
\end{align}
Physically, $\nu$ represents the fraction of the $s=1$ state during one turnover cycle of the enzymatic reaction. 
It depends only on the substrate concentration and the bare rate constants.
In the following analyses, we vary this state parameter $\nu$ to investigate the shear viscosity of enzyme solutions.
Some numerical estimates of $\nu$ are given in the end of this section.

We discuss here the validity of using a single-enzyme kinetics.
In our model, we have assumed that the concentration of enzymes is small enough so that hydrodynamic interactions between 
enzymes are negligible~\cite{Doibook}.
Such a dilute condition corresponds to having only a single enzyme in the system, leading to a renewal process~\cite{Lu1998}.
In the renewal process, the probability distribution function is identically and independently distributed~\cite{Saha2011}.
This means that in every turnover cycle, waiting times follow the same probability distribution, and hence these times can be uniquely 
determine as shown in Eq.~(\ref{P1}).

For systems containing mesoscopic numbers of enzymes, however, stochasticity in enzymatic reactions plays more 
important roles as discussed in Refs.~\cite{Grima2009,Saha2011}.
Enzyme stochasticity leads to non-renewal processes and causes breakdown of the Michaelis-Menten equation in steady state~\cite{Grima2009,Saha2011}.
Since the waiting time distributions depends on the number of enzymes for non-renewal processes, 
one needs to derive master equations for waiting time distributions when a solution of multiple enzymes 
is considered~\cite{Saha2011}.
This is beyond the scope of the present work.

\subsection{Viscosity of two-state dimer solutions}

To calculate the shear viscosity of a two-state enzyme solution, we introduce the following notations: 
$\kappa=K_1/K_0$, 
$\lambda=\ell_1/\ell_0$, and 
$\lambda^\ast=\ell^\ast/\ell_0=(1+\kappa\lambda)/(1+\kappa)$, where 
$\ell^\ast=(K_0\ell_0+K_1\ell_1)/(K_0+K_1)$ is the effective natural length for a dimer 
 in the $s=1$ state.
In Appendix~\ref{app:eta_a}, we show that the viscosity of a two-state enzyme solution is given by
\begin{align}
\eta_{\rm e}(\nu,\epsilon, \kappa, \lambda)=\eta_{\rm 0} + \left(  \eta_1- \eta_{\rm 0}\right)
\frac{z \nu}{1+z \nu},
\label{eq:eta_E}
\end{align}
where the quantity $\eta_1$ ($\eta_0$) corresponds to the viscosity when all the enzymes are in the $s=1$ ($s=0$) state
\begin{align}
\frac{\eta_1(\epsilon, \kappa, \lambda)}{G\tau}=
\frac{2\epsilon}{3} \frac{g_4\left(\epsilon(1+\kappa), \lambda^\ast \right)}{g_2\left(\epsilon(1+\kappa), \lambda^\ast \right)},
\label{eq:eta_1}
\end{align}
and
\begin{align}
z(\epsilon,\kappa, \lambda)=\exp\left[-\frac{\epsilon\kappa}{1+\kappa}(\lambda-1)^2\right]
\frac{g_2\left(\epsilon(1+\kappa), \lambda^\ast \right)}
{g_2(\epsilon, 1)}.
\label{eq:Z_10}
\end{align}
See also Eq.~(\ref{eq:eta_F}) for the Fraenkel dimer viscosity $\eta_{\rm 0}(\epsilon)$.
In the above, $g_m(p,q)$ is given by an integral
\begin{align}
g_m(p,q)=\int_0^\infty dr\, r^m e^{-p(r-q)^2},
\end{align} 
and its explicit expression is obtained in Appendix~\ref{app:eta_a}
[see Eq.~(\ref{eq:ap_g})].
Specifically, the functions $g_2(p, q)$ ($m=2$) and $g_4(p, q)$ ($m=4$) are given by
\begin{align}
g_2(p, q)=\frac{q}{2p}e^{-pq^2}
+\frac{\sqrt{\pi}(1+2pq^2)[1+{\rm erf}(\sqrt{p}q)]}{4p^{3/2}},
\end{align}

\begin{figure}[tbh]
\begin{center}
\includegraphics[scale=0.6]{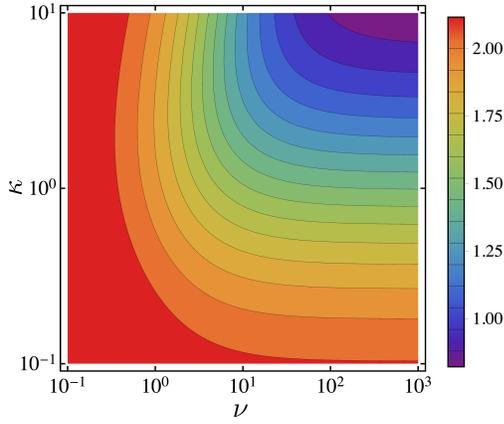}
\end{center}
\caption{
(Color online)
Contour plot of $\eta_{\rm e}/(G\tau)$ as a function of the parameters $\nu=c_{\rm S}/K_{\rm M}$
[see Eq.~(\ref{eq:nu})] and $\kappa=K_1/K_0$ for $\epsilon=K_0\ell_0^2/(2k_{\rm B}T)=1$ and 
$\lambda=\ell_1/\ell_0=1$.
}
\label{fig:nu_mu_etaE_chi1}
\end{figure}

\begin{figure}[tbh]
\begin{center}
\includegraphics[scale=0.6]{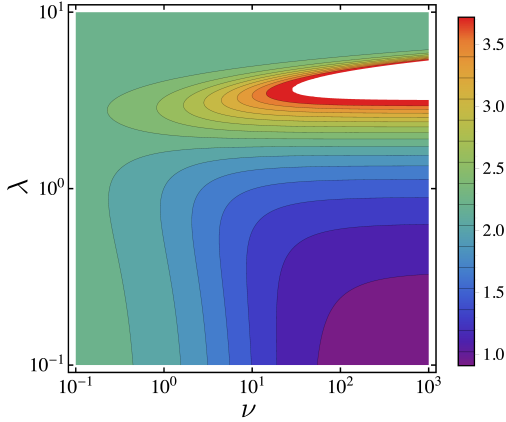}
\end{center}
\caption{
(Color online)
Contour plot of $\eta_{\rm e}/(G\tau)$ as a function of the parameters $\nu=c_{\rm S}/K_{\rm M}$
[see Eq.~(\ref{eq:nu})] and $\lambda=\ell_1/\ell_0$ for $\epsilon=K_0\ell_0^2/(2k_{\rm B}T)=1$ and 
$\kappa=K_1/K_0=1$.
The white region corresponds to larger absolute values of $\eta_{\rm e}$.
}
\label{fig:nu_chi_etaE_mu1}
\end{figure}

and
\begin{align}
g_4(p, q)&=\frac{q(5+2pq^2)}{4p^2}e^{-pq^2}\nonumber\\
&+\frac{\sqrt{\pi}(3+12pq^2+4p^2q^4)[1+{\rm erf}(\sqrt{p}q)]}{8p^{5/2}},
\label{eq:g4}
\end{align}
respectively.
Equations~(\ref{eq:eta_E})--(\ref{eq:g4}) for the viscosity are the main result of this work.

In Eq.~(\ref{eq:Z_10}), the factor $\epsilon\kappa/(1+\kappa)(\lambda-1)^2$ in the exponential function 
corresponds to the dimensionless energy difference, $U(\ell^\ast,1)-U(\ell_0,0)$, between the two equilibrium states of a two-state dimer with $\ell_0$ and $\ell^\ast$, as shown in Fig.~\ref{fig:cycle}(b).
Although only the bare reaction rates are taken into account, the above energy difference naturally 
emerges by defining the weighted distribution function as in Eq.~(\ref{eq:psi_r}).

\begin{figure}[tbh]
\begin{center}
\includegraphics[scale=0.35]{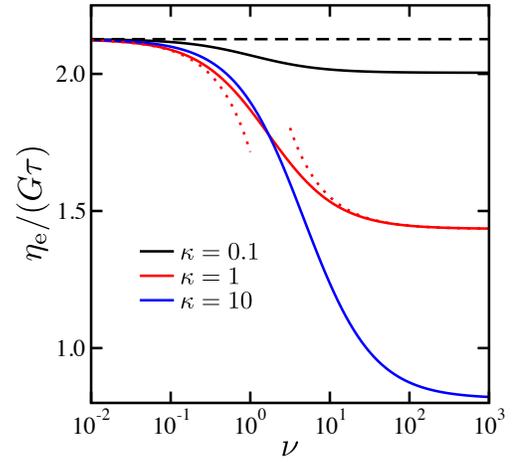}
\end{center}
\caption{
(Color online)
Plot of $\eta_{\rm e}/(G\tau)$ as a function of the parameter $\nu$ for 
$\kappa=K_1/K_0=0.1,1$ and $10$.
The other parameter values are $\epsilon=K_0\ell_0^2/(2k_{\rm B}T)=1$ and 
$\lambda=\ell_1/\ell_0=1$.
The black dashed line represents $\eta_{\rm 0}$ in Eq.~(\ref{eq:eta_F}).
The red dotted lines represent the two limiting expressions in Eq.~(\ref{eq:C1C2}) for $\kappa=1$.
\label{fig:nu_mu_etaE_chi1_2D}
}
\end{figure}

\begin{figure}[tbh]
\begin{center}
\includegraphics[scale=0.35]{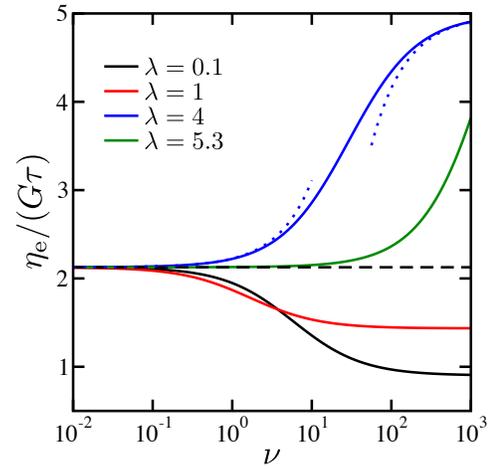}
\end{center}
\caption{
(Color online)
Plot of $\eta_{\rm e}/(G\tau)$ as a function of the parameter $\nu$ for 
$\lambda=\ell_1/\ell_0=0.1,1,4$ and $5.3$.
The other parameter values are $\epsilon=K_0\ell_0^2/(2k_{\rm B}T)=1$ and 
$\kappa=K_1/K_0=1$.
The black dashed line represents $\eta_0$ in Eq.~(\ref{eq:eta_F}).
The blue dotted lines represent the two limiting expressions in Eq.~(\ref{eq:C1C2}) for $\lambda=4$.
\label{fig:nu_chi_etaE_mu1_2D}
}
\end{figure}

When $\nu=0$, $\eta_{\rm e}$ of Eq.~(\ref{eq:eta_E}) simply reduces to $\eta_0$, the viscosity of the Fraenkel dimer 
solution [$s=0$, see Eq.~(\ref{eq:eta_F})].
For $\nu\neq0$, the enzyme solution viscosity $\eta_{\rm e}$ is determined by the ratio between the two viscosities $\eta_{\rm 0}$ and $\eta_1$.
Due to the factor $z$, however, $\eta_{\rm e}$ also depends on the energy difference between the two states of the enzyme.
This effect causes a non-monotonic behavior of the viscosity as we will show later.

Before analyzing the behavior of $\eta_{\rm e}$, we estimate typical values of 
$\epsilon=K_0\ell_0^2/(2k_{\rm B}T)$.
The enzymes size can be taken as $\ell_0\approx10$\,nm~\cite{Albertsbook}.
Moreover, considering typical forces, $1$\,pN, generated by a two-state dimer with size $\ell_0$, we 
estimate the spring constant as $K_0\approx10^{-4}$\,N/m~\cite{Mikhailov2015}.
Using these values and $k_{\rm B}T\approx4\times10^{-21}$\,J in physiological conditions, we obtain $\epsilon \approx 1$. 
Hence, we fix the $\epsilon$ value hereafter to $\epsilon = 1$.

In Fig.~\ref{fig:nu_mu_etaE_chi1}, we present the contour plot of the rescaled viscosity due to two-state dimers, 
$\eta_{\rm e}/(G \tau)$,  as a function of $\nu$ and $\kappa$ for $\epsilon=\lambda=1$.
One can see that $\eta_{\rm e}$ becomes smaller for large $\nu$ and $\kappa$, implying that the viscosity decreases when enzymatic reactions occur more frequently and substrates are stiffer (large $K_1$).
Notice that stiff dimers lead to a decrease of $\eta_{\rm e}$ because its stiffness suppresses the enzyme size fluctuation.
In Fig.~\ref{fig:nu_chi_etaE_mu1}, we plot the rescaled viscosity, $\eta_{\rm e}/(G \tau)$, as a function of $\nu$ and $\lambda$ for $\epsilon=\kappa=1$.
Here we see a non-monotonic behavior of the viscosity in $\lambda$ characterized by a peak around $\lambda\approx 3.2$.
Note that for larger $\lambda$ values, $\eta_{\rm e}$ becomes independent of $\nu$.

To see more detailed behavior, we plot in Fig.~\ref{fig:nu_mu_etaE_chi1_2D} the 
rescaled viscosity, $\eta_{\rm e}/(G \tau)$, as a function of $\nu$ for $\kappa=0.1$, $1$ and $10$, while keeping $\epsilon=\lambda=1$.
The dashed line corresponds to the constant viscosity for a Fraenkel dimer solution, i.e.,
$\eta_{\rm 0}/(G \tau) \approx 2.13$.
We see that $\eta_{\rm e}$ decreases with increasing $\nu$ for all the $\kappa$ values. 
The decrease of $\eta_{\rm e}$ is more enhanced for larger $\kappa$ values.

In Fig.~\ref{fig:nu_chi_etaE_mu1_2D}, we plot $\eta_{\rm e}$ as a function of $\nu$ for 
$\lambda=0.1$, $1$, $4$ and $5.3$, while keeping $\epsilon=\kappa=1$.
We see that $\eta_{\rm e}$ shows both increasing and decreasing dependency as a function of $\nu$ depending on the value of $\lambda$.
When $\lambda=0.1$, $1$, and $4$, the viscosity $\eta_{\rm e}$ increases with $\lambda$, 
reflecting the fact that larger enzymes lead to higher viscosity.
For larger $\lambda$ such as $\lambda=5.3$, however, $\eta_{\rm e}$ becomes smaller, and as $\lambda$ 
is further increased, the viscosity approaches the value of $\eta_{\rm 0}$ as indicated by the dashed line.
In this limit, both Fraenkel dimer solutions and two-state enzyme solutions exhibit the same viscosity even when $\nu$ is very large.

We discuss now the non-monotonic behavior of $\eta_{\rm e}$ that is seen in Fig.~\ref{fig:nu_chi_etaE_mu1_2D}.
Such a behavior occurs because $z$ in Eq.~(\ref{eq:Z_10}) increases for smaller $\lambda$, but strongly 
decreases for larger $\lambda$ due to the Gaussian function of Eq.~(\ref{eq:Z_10}).
The factor $\epsilon\kappa(\lambda-1)^2/(1+\kappa)$ in the Gaussian function corresponds to the rescaled energy 
difference between the $s=0$ and $s=1$ states.
Hence, it can be regarded as an Arrhenius' equation that determines the transition rate from the $s=0$ to $s=1$ state.

\begin{figure}[tbh]
\begin{center}
\includegraphics[scale=0.6]{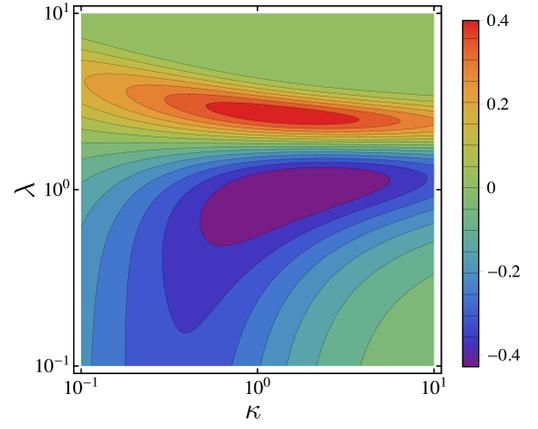}
\end{center}
\caption{
(Color online)
Contour plot of $C_1/(G\tau)$ [see Eq.~(\ref{eq:C1C2})] as a function of $\kappa=K_1/K_0$ and 
$\lambda=\ell_1/\ell_0$ for $\epsilon=K_0\ell_0^2/(2k_{\rm B}T)=1$ under the condition $\nu\ll1$.
The quantity $C_1$ changes its sign from negative to positive around $\lambda\approx2$.
}
\label{fig:nu_chi_F1}
\end{figure}

\begin{figure}[tbh]
\begin{center}
\includegraphics[scale=0.6]{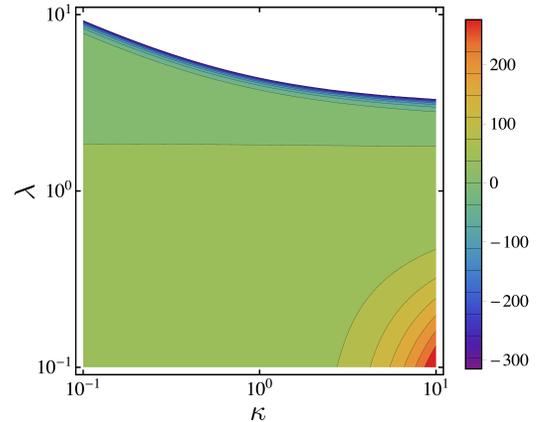}
\end{center}
\caption{
(Color online)
Contour plot of $C_2/(G\tau)$ [see Eq.~(\ref{eq:C1C2})] as a function of $\kappa=K_1/K_0$ and 
$\lambda=\ell_1/\ell_0$ for $\epsilon=K_0\ell_0^2/(2k_{\rm B}T)=1$ under the condition $\nu\gg1$.
The quantity $C_2$ changes its sign from positive to negative around $\lambda\approx2$.
The white region corresponds to larger absolute values of $C_2$.
}
\label{fig:nu_chi_F2}
\end{figure}

\subsection{Limiting expressions}

Next, we present the limiting expressions of $\eta_{\rm e}$ for small and large values of the $\nu$ parameter, $\nu \ll 1$ and $\nu \gg 1$.
The viscosity of two-state dimer solution in Eq.~(\ref{eq:eta_E}) becomes
\begin{align}
\eta_{\rm e}(\nu, \epsilon, \kappa, \lambda) \approx \left\{ \begin{array}{lr}{\displaystyle \eta_{\rm 0} +  C_1 \nu }, & {~~~\nu\ll1} \\[2.0ex]
 {\displaystyle \eta_1 + \frac{C_2}{\nu} }, & {~~~\nu\gg1}
\end{array} \right.
\label{eq:C1C2}
\end{align}
where $C_1(\epsilon, \kappa, \lambda)=(\eta_1-\eta_{\rm 0})z$ and $C_2(\epsilon, \kappa, \lambda)=(\eta_{\rm 0}-\eta_1)/z$.
In Figs.~\ref{fig:nu_mu_etaE_chi1_2D} and \ref{fig:nu_chi_etaE_mu1_2D}, we have plotted 
the above limits by the red (for $\kappa=1$) and blue (for $\lambda=4$) 
dotted line, respectively.

In Fig.~\ref{fig:nu_chi_F1}, we study the $\nu\ll1$ behavior and plot the coefficient $C_1=(\eta_1-\eta_{\rm 0})z$ of $\nu$
in Eq.~(\ref{eq:C1C2}) as a function of $\kappa$ and $\lambda$ for $\epsilon=1$.
The behavior of $C_1$ is non-monotonic, having a minimum and a maximum around 
$(\kappa,\lambda)\approx (1, 1)$ and $(\kappa,\lambda)\approx(1, 2.5)$, respectively.
The quantity $C_1$ vanishes for large $\lambda$ values, because the Gaussian function in $z$, Eq.~(\ref{eq:Z_10}), dominates over the viscosity difference, $\eta_0-\eta_1$.
Notice that $C_1$ changes its sign from negative to positive around $\lambda\approx 2$, 
where the switching from decreasing to increasing behavior of $\eta_{\rm e}$ as a function of $\nu$ occurs.

In Fig.~\ref{fig:nu_chi_F2}, we study the $\nu\gg1$ behavior and plot the coefficient $C_2=(\eta_{\rm 0}-\eta_1)/z$ 
of $\nu^{-1}$ in Eq.~(\ref{eq:C1C2}) as a function of $\kappa$ and $\lambda$ when $\epsilon=1$.
Here $C_2$ exhibits a monotonic behavior in $\kappa$ and $\lambda$, and changes its sign from positive to negative 
around $\lambda\approx 2$.
Since $\eta_{\rm e}$ is inversely proportional to $\nu$ in Eq.~(\ref{eq:C1C2}), positive $C_2$ leads to a 
decreasing behavior of $\eta_{\rm e}$, whereas negative $C_2$ results in an increasing behavior.

\subsection{Numerical estimates}

To end this section, we give some numerical estimates of the parameter 
$\nu=c_{\rm S}/K_{\rm M}$ in Eq.~(\ref{eq:nu}).
The experimentally accessible substrate concentration is 
$10^{-6}\,{\rm M}<c_{\rm S}<10^{-3}$\,M~\cite{Zhao2017,Jee2018}.
On the other hand, the value of the Michaelis constant $K_{\rm M}$ differs between fast 
and slow enzymes.
For fast enzymes, such as urease and catalase, it is given by 
$K_{\rm M} \approx 10^{-3}$\,M~\cite{Riedel2015,Jee2018}.
For slow enzymes, such as aldolase and adenylate kinase, it is  
$K_{\rm M}\approx10^{-6}$\,M~\cite{Illien2017_2,Aviram2018}.
Hence, the $\nu$ range is estimated as $10^{-3}<\nu<1$ and $1<\nu<10^3$, respectively, for fast and slow enzymes.
These estimates imply that the limiting expressions derived for $\nu\ll1$ and $\nu\gg1$ in Eq.~(\ref{eq:C1C2}) correspond to these two types of enzymes for 
$c_{\rm S}<10^{-4}$\,M and $c_{\rm S}>10^{-5}$\,M, respectively.

Next we discuss the values of $\kappa$ and $\lambda$ in order to estimate the viscosity $\eta_{\rm e}$ 
for typical physiological conditions.
Since an enzyme consists of a large complex of macromolecules, the size of substrate molecules is typically 
smaller than that of enzymes~\cite{Albertsbook}.
Due to this size difference, the condition $\lambda <1$ holds generally.
Non-covalent bonds, such as hydrogen bonds, van der Waals attractions and 
hydrophobic forces, are responsible for the formation of macromolecular assemblies.
On the other hand, covalent bonds are responsible for the formation of substrate molecules.
Then, the molecular flexibilities for the substrates compared with the enzymes are different, which leads to the condition $\kappa > 1$.

From the above argument, we choose $\lambda=0.1$ and $\kappa = 10$.
Using these values and setting $\epsilon=1$, we obtain $\eta_{\rm e}/(G\tau) \approx 2.11$ and 
$\eta_{\rm e}/(G\tau) \approx 0.39$ for fast and slow enzymes, respectively, assuming that the maximum 
substrate concentration $c_{\rm S}=10^{-3}$\,M is attained. 
Since $\eta_{\rm 0}/(G\tau) \approx2.13$ for $\epsilon=1$, the difference between the enzyme solution with substrates $\eta_{\rm e}$ and that without substates $\eta_{\rm 0}$ is negligible for fast enzymes, whereas the viscosity 
$\eta_{\rm e}$ is approximately five times smaller than $\eta_{\rm 0}$ for slow enzymes.

\section{Discussion and conclusion}
\label{sec:discussion}

In this paper, we have investigated the viscosity of dilute two-state enzyme solutions under steady shear flow.
We have obtained the shear viscosity by taking into account the enzyme conformational changes in a solution with 
a supply of substrates.
The waiting times, which correspond to the respective conformations of the enzyme, are connected to 
the reaction rates in the enzymatic cycle by using the single enzyme kinetics~\cite{Lu1998}.
In our approach, the two-state dimer model~\cite{Mikhailov2015,Flechsig2019,Hosaka2019} 
and the polymer dimer model~\cite{Fraenkel1952,Birdbook,Doibook} are combined.

When the enzyme has the same structural properties as the substrate, the shear viscosity  
decreases as the substrate concentration becomes higher (see Fig.~\ref{fig:nu_mu_etaE_chi1_2D}).
For a substrate larger than the enzyme, the viscosity increases with substrate concentrations 
(see Fig.~\ref{fig:nu_chi_etaE_mu1_2D}).
When the substrate is large enough, however, the viscosity reduces to that of a Fraenkel dimer solution.
Furthermore, we have obtained the limiting expressions of the viscosity for fast and slow enzymes 
[see Eq.~(\ref{eq:C1C2})].
For slow enzymes, the coefficient shows only a monotonic behavior.
For fast enzymes, on the other hand, the coefficient of the substrate concentration exhibits a non-monotonic 
behavior as functions of  the stiffness and size of the substrate.

Next, we comment on the connection between the viscosity of a two-state dimer solution and the diffusion coefficient of a tracer particle in such a solution.
By following the discussion in Refs.~\cite{Oppenheimer2009,Oppenheimer2010}, the diffusion coefficient of a passive spherical particle of radius $R$ can be given by Einstein's relation
\begin{align}
D_{\rm e}=\frac{k_{\rm B}T}{6\pi(\eta_{\rm s}+\eta_{\rm e})R},
\end{align}
where we have assumed $R\gg\ell_0$.
In terms of the enzyme volume fraction $\phi=4\pi (\ell_0/2)^3n/3$, $D_{\rm e}$ can be expanded up 
to first order in $\phi$ as 
\begin{align}
D_{\rm e} \approx \frac{k_{\rm B}T}{6\pi\eta_{\rm s}R}
\left( 1 - \frac{9a\eta_{\rm e}}{2\ell_0G\tau} \phi \right).
\label{eq:Dact}
\end{align}
Hence, the relative change of the diffusion coefficient with respect to that of a Fraenkel dimer solution 
(denoted by $D_0$) is 
\begin{align}
\delta D= D_{\rm e}-D_{\rm 0}=
\frac{3k_{\rm B}T}{4\pi\eta_{\rm s}R}\frac{a(\eta_{\rm 0}-\eta_{\rm e})}{\ell_0G\tau}\phi.
\label{eq:deltaD}
\end{align}

Since $\eta_{\rm 0}>\eta_{\rm e}$ holds for both fast and slow enzymes as estimated before, catalytic 
enzymes give rise to the diffusion enhancement under physiological conditions.
Moreover, we see that $\delta D$ increases as $c_{\rm S}$ is increased in the limits of fast and slow enzymes 
(see Figs.~\ref{fig:nu_chi_F1} and \ref{fig:nu_chi_F2}).
This behavior qualitatively agrees with experiments for both tracers and enzymes~\cite{Muddana2010,Zhao2017,Xu2019}.
More specifically, using values such as $c_{\rm S}=10^{-3}$\,M, $a/\ell_0=0.2$, $\phi=0.1$, we obtain that
the diffusion increases for slow enzymes as $\delta D/D_{\rm 0}\approx0.15$.
In existing experiments, however, $\phi$ is typically of the order of $10^{-5}$, and hence experimental measurements using higher $c_{\rm E}$ concentration are needed for a more accurately checking of the validity of our model.

Here we discuss how the obtained viscosity is modified by hydrodynamic effects that have been neglected so far.
In the presence of hydrodynamic interactions, the equation of motion, Eq.~(\ref{eq:velocity}), can be rewritten as~\cite{Bird1971}
\begin{align}
\frac{\partial r_\alpha}{\partial t}=(\delta_{\alpha\beta}-\zeta G_{\alpha\beta})\left(
\frac{2}{\zeta}f_\beta
-\frac{2k_{\rm B}T}{\zeta}\frac{\partial \ln\psi}{\partial r_\beta}
\right)
+d_{\alpha\beta}r_\beta,
\label{eq:Oseen}
\end{align}
where $G_{\alpha\beta}(r)=\left(\delta_{\alpha\beta}+r_\alpha r_\beta/r^2\right)/(8\pi\eta_{\rm s} r)$ is the 
hydrodynamic Oseen tensor~\cite{Doibook2}.
If we assume all orientations to be equally probable, an equilibrium-averaged hydrodynamic interaction can be defined by 
taking the average of $G_{\alpha\beta}(r)$ over all orientations~\cite{Warner1972}
\begin{align}
h=
\frac{1}{3}{\rm Tr} \left(\frac{\int d\textbf{r}\, \psi(r)G_{\alpha\beta}(r)}{\int d\textbf{r}\, \psi(r)}\right),
\label{eq:h}
\end{align}
where Tr denotes the trace operation.
This is called the pre-averaging approximation~\cite{Doibook2}.
Then, the equation of motion can be approximated as
\begin{align}
\frac{\partial r_\alpha}{\partial t}& \approx
\frac{2(1-\zeta h)}{\zeta}
\left(
f_\alpha
-k_{\rm B}T\frac{\partial \ln\psi}{\partial r_\alpha}
\right)
+d_{\alpha\beta}r_\beta.
\label{eq:Oseen2}
\end{align}

Comparing Eqs.~(\ref{eq:velocity}) and (\ref{eq:Oseen2}), one finds that the change over from negligible hydrodynamic interactions 
to equilibrium-averaged ones can be accomplished by replacing $\zeta$ with $\zeta/(1-\zeta h)$.
Hence, for a single-state dimer as in Eq.~(\ref{eq:eta_simple}), the hydrodynamic interaction modifies the viscosity by 
a factor of $1/(1-\zeta h)$.
In Appendix~\ref{app:hydro}, we derive $h$ for the Fraenkel dimer model.
When $a/\ell_0=0.2$ and $\epsilon=1$, for example, we find that the viscosity is about 20\% larger as compared 
to the negligible hydrodynamic case.
For the two-state dimers, hydrodynamic effects do not affect the $\nu$-dependence of $\eta_{\rm e}$ although
some geometrical factors such as $\kappa$ and $\lambda$ can enter in $h$.

In this study, we have assumed that the distribution functions do not depend on shear flow 
[see Eqs.~(\ref{passive}) and (\ref{eq:psi})].
Here we discuss how these distribution functions are modified by an external flow and the regime where the flow does not affect the distributions as assumed in this paper.
For a steady-state homogeneous potential flow, Eq.~(\ref{eq:FP}) has an analytical solution~\cite{Birdbook}
\begin{align}
\psi(r) =C^\prime \exp\left[ -\frac{U(r)}{k_{\rm B}T}\right]\exp\left[ \frac{\zeta}{k_{\rm B}T}r_\alpha d_{\alpha\beta}r_\beta
\right],
\end{align} 
where $C^\prime$ is the normalization  constant.

For a simple shear flow characterized by a shear rate $\dot{\gamma}$, the distribution function becomes
\begin{align}
\psi(r,\theta,\phi, \dot{\gamma}) =C^\prime \exp\left[ -\frac{U(r)}{k_{\rm B}T}\right]
\exp\left[ \frac{\zeta r^2 \dot{\gamma}}{2k_{\rm B}T}\sin^2\theta\sin2\phi
\right],
\label{eq:psi_gamma}
\end{align} 
where $r_x=r\sin\theta\cos\phi$ and $r_y=r\sin\theta\sin\phi$.
When the length of a dimer is $r=\ell_0$, the characteristic relaxation time is given by $\zeta\ell_0^2/(k_{\rm B}T)$~\cite{Fraenkel1952}.
Hence, the shear flow does not affect the distribution functions when $\zeta\ell_0^2/(k_{\rm B}T)\dot{\gamma}\ll1$.

We have assumed that the transition time spent from one enzymatic species to another is much smaller than the waiting time, i.e., $\tau/W_s\ll1$.
Here, we consider the general case of arbitrary waiting time.
Because the total times in state $s=0$ and $s=1$ are given by $W_0+\tau$ and $W_1+\tau_1$, respectively, 
the modified parameter $\nu$ becomes
\begin{align}
\nu=\frac{k_1(1+k_{\rm cat}\tau_1)c_{\rm S}}
{k_{-1}+k_{\rm cat}(1+k_1\tau c_{\rm S} )},
\label{eq:nuprime}
\end{align}
where $\tau=\zeta/(4K_0)$ as before and $\tau_1=\zeta/(4K_1)$.
Since the reverse reaction rate $k_{-1}$ is negligible in general but may have a finite value, we set 
it to be a constant.
There are only four relevant time scales, namely, $k_{\rm cat}^{-1}$, $(k_1 c_{\rm S})^{-1}$, $\tau$, and $\tau_1$,
and Eq.~(\ref{eq:nuprime}) has four limiting expressions.
When the transition rates are vanishingly small, the modified parameter coincides with $\nu$ in Eq.~(\ref{eq:nu}) as it should.
For the two intermediate regimes, Eq.~(\ref{eq:nuprime}) shows linear and inverse dependences on the transition time.
When the transition time is infinitely large, we have $\nu \sim \kappa^{-1}$, indicating that 
the transition dynamics is governed only by the relative stiffness between the enzyme and substrate.

The transition rates can depend on $\kappa$ and/or  $\lambda$ for general enzymatic solutions although these effects were not considered in this work.
Using Kramers' reaction-rate theory~\cite{Haenggi1990}, Aviram \textit{et al.}~\cite{Aviram2018} obtained free-energy profiles of enzymes by 
experimentally measuring the transition rates.
In the presence of such an effect, the enzyme solution viscosity may exhibit more complicated dependences on 
$\kappa$ and/or $\lambda$.
Finally, we have assumed that the viscosity due to enzymes does not depend on the shear rate.
Since the dimer model with finite natural lengths predicts a viscosity that depends on the shear 
rate~\cite{Birdbook,Bird1997}, one can extend the present model to a non-Newtonian enzymatic fluid.

\acknowledgements

We thank R.\ M.\ Adar, Y.\ Avni, K.\ K.\ Dey, V.\ D\'{e}mery, M.\ Doi, T.\ Kato, A.\ S.\ Mikhailov, and K.\ Yasuda for fruitful discussions 
and helpful suggestions.
Y.H.\ acknowledges support by a Grant-in-Aid for JSPS Fellows (Grant No.\ 19J20271) from the Japan Society 
for the Promotion of Science (JSPS). 
Y.H.\ also thanks the hospitality of Tel Aviv University,  where part of this research was conducted under the 
TMU-TAU co-tutorial program. 
S.K.\ acknowledges the support by Grant-in-Aid for Scientific Research (C) (Grant No.\ 18K03567 and 
No.\ 19K03765) from the JSPS.
D.A.\ acknowledges support from the Israel Science Foundation (ISF) under grant no.\ 213/19.

\appendix
\section{Probability distribution function for multiple-state enzymes}
\label{app:general_psi}

In this Appendix, we generalize the dimer-enzyme into a $N$-mer one.
We derive the probability distribution function for a single enzyme that  has multiple intermediate 
states in catalytic chemical reactions.
We consider the following cascade reaction containing $N$ intermediate substrate-enzyme 
complexes:
\begin {align}
{\rm E} + {\rm S} 
\overset {k_1}{\underset {k_{-1}}{\rightleftarrows }} {\rm (ES)_1} 
\overset {k_2}{\underset {k_{-2}}{\rightleftarrows }} {} 
\cdots
{\rm (ES)}_s
\cdots
\overset {k_N}{\underset {k_{-N}}{\rightleftarrows }} {\rm (ES)}_N
\xrightarrow {k_{\rm cat}}{\rm E}_\ast + {\rm P}.
\label{eq:cascade}
\end{align}
Here $({\rm ES})_s$ denotes the $s$-th intermediate complex in the reaction, and $k_s$ and $k_{-s}$ 
are the forward and backward reaction rates to the states $s$ and $s-1$, respectively.
At the final step, the complex  is irreversibly converted to an enzyme and a product with the reaction rate 
$k_{\rm cat}$.
The enzyme after the catalysis is denoted by ${\rm E}_\ast$.

Since we assume that a substrate having the energy $E_s$ binds to $({\rm ES})_{s-1}$ with the reaction 
rate $k_s$, the energy of an enzyme in the state $s$ can be written as 
\begin{align}
U(r,s)&=E_0 + \sum_{s'=1}^{s}E_{s'},
\end{align}
where $E_0$ is the energy of the free enzyme.
Then, the waiting time-weighted distribution functions is given by 
\begin{align}
\psi_{N}(r)
&= \frac{\sum_{s=0}^N W_s e^{-\beta U(r,s)}}
{\sum_{s=0}^N W_s\int d\mathbf{r}\, e^{-\beta U(r,s)}}.
\label{eq:general_dist}
\end{align}
Here $W_s$ is the waiting time in the state $s$, which is defined in Eq.~(\ref{eq:p_i}).

In order to obtain the viscosity of dimer solutions using Eq.~(\ref{eq:eta_simple}), 
we need to calculate the second moment $\langle r_y^2 \rangle$.
In general, the average of any function $f(\mathbf{r})$ over the distribution function, Eq.~(\ref{eq:general_dist}), can be written as
\begin{align}
& \langle f(\mathbf{r}) \rangle_{N} =\langle f(\mathbf{r}) \rangle_0  \nonumber  \\
&+ \sum_{s=1}^N \left[ \langle f(\mathbf{r}) \rangle_s-\langle f(\mathbf{r}) \rangle_0\right]
\frac{z_{s0} w_{s0}}{1+\sum_{s'=1}^N z_{s'0} w_{s'0}},
\label{eq:f_r}
\end{align}
where $\langle f(\mathbf{r}) \rangle_s$ denotes the average of $f(\mathbf{r})$ over all configurations in the state $s$
\begin{align}
\langle f(\mathbf{r}) \rangle_s &= \frac{\int d\mathbf{r}\, f(\mathbf{r})e^{- \beta U(r,s)}}
{\int d\mathbf{r}\, e^{- \beta U(r,s)}},
\end{align}
while $z_{ss'}$ and $w_{ss'}$ are defined by 
\begin{align}
z_{ss'}=\frac{\int d\mathbf{r}\, e^{- \beta U(r,s)}}
{\int d\mathbf{r}\, e^{- \beta U(r,s')}},~~~~~
w_{ss'}=\frac{\int_0^\infty dt\, p_s(t)}{\int_0^\infty dt\, p_{s'}(t)}.
\label{Zij} 
\end{align}
Notice that the quantity $z$ in Eq.~(\ref{eq:Z_10}) corresponds to $z_{10}$ in the above notation.

\section{Michaelis-Menten kinetics and single enzyme kinetics}
\label{app:MM_single}

In this Appendix,  we briefly review the Michaelis-Menten kinetics~\cite{Michaelis1913} and the single-enzyme kinetics.
In the two-state dimer model, the cascade reaction in Eq.~(\ref{eq:cascade}) reduces to the 
Michaelis-Menten reaction [see Eq.~(\ref{eq:MM})].
In the ensemble of enzymatic experiments, the corresponding kinetic equations become
\begin{align}
\frac{d c_{\rm E} }{dt} &=  k_{-1} c_{\rm ES} - k_1 c_{\rm E} c_{\rm S} , \nonumber \\
\frac{d c_{\rm ES} }{dt}  &=  k_{1} c_{\rm E} c_{\rm S} - (k_{-1}+k_{\rm cat}) c_{\rm ES}, \nonumber \\
\frac{d c_{\rm P} }{dt} &= k_{\rm cat} c_{\rm ES},
\end{align}
where $c_{\rm E}$ and $c_{\rm S}$ were defined before, whereas $c_{\rm ES}$ and $c_{\rm P}$ are the 
concentrations of substrate-enzyme complex and product, respectively.
By replacing the concentrations of the chemical species with the probability distributions, 
we obtain the kinetic equations for a single enzyme as in Eq.~(\ref{eq:dp01e0}).
In the steady sate, $d c_{\rm ES}/dt=0$, the enzymatic velocity is given by 
\begin{align}
V = \frac{d c_{\rm P}}{dt} = \frac{V_{\rm max} c_{\rm S}}{K_{\rm M}+c_{\rm S}},
\label{eq:MMeq}
\end{align}
where $V_{\rm max}=k_{\rm cat}( c_{\rm E} + c_{\rm ES})$ is the maximum enzymatic velocity and 
$K_{\rm M}=(k_{-1}+k_{\rm cat})/k_1$ is the Michaelis constant defined in Eq.~(\ref{eq:MMconst}).

For a single-enzyme, the corresponding reaction velocity can be obtained from the inverse of the total waiting time during one catalytic cycle.
With the use of Eq.~(\ref{P1}), this velocity becomes
\begin{align}
\frac{1}{W}=\frac{1}{W_0+W_1} = \frac{k_{\rm cat} c_{\rm S}}{K_{\rm M}+c_{\rm S}},
\label{eq:singleMMeq}
\end{align}
which is termed the single-molecule Michaelis-Menten equation~\cite{Kou2005}.
Comparison of Eqs.~(\ref{eq:MMeq}) and (\ref{eq:singleMMeq}) yields the relation 
\begin{align}
\frac{V}{ c_{\rm E}+ c_{\rm ES} }=\frac{1}{W}.
\end{align}
This relation originates from the equivalence between the average over a single molecule's long-time trace and that over a large ensemble of identical molecules, i.e., the ergodicity~\cite{Kou2005,English2006}.

\section{Derivation of $\eta_{\rm e}$}
\label{app:eta_a}

In this Appendix, we present the derivation of $\eta_{\rm e}$ in Eq.~(\ref{eq:eta_E}).
Using Eq.~(\ref{eq:psi_r}), we calculate $\langle r_y^2 \rangle$ in Eq.~(\ref{eq:eta_simple}) as 
\begin{align}
\eta_{\rm e} & = \frac{n\zeta}{4}
\frac{ \int d\mathbf{r}\, \left[ r_y^2e^{-\beta U(r,0)} + \nu  r_y^2e^{-\beta U(r,1)} \right] }
{\int d\mathbf{r}\, \left[ e^{-\beta U(r,0)} + \nu e^{-\beta U(r,1)} \right] }.
\end{align}
With the use of Eq.~(\ref{eq:f_r}) for $N=1$, we obtain
\begin{align}
\eta_{\rm e} &=  \frac{n\zeta}{4}\left( \langle r_y^2 \rangle_0 + \left[ \langle r_y^2 \rangle_1-\langle r_y^2 \rangle_0\right] \frac{z\nu}{1+z\nu}
\right).
\end{align} 
Since $n\zeta \langle r_y^2 \rangle_0/4=\eta_{\rm 0}$ and 
$n\zeta\langle r_y^2 \rangle_1/4 =\eta_1$, we obtain Eq.~(\ref{eq:eta_E}).
The viscosity of a Fraenkel dimer solution $\eta_{\rm 0}$ is given by Eq.~(\ref{eq:eta_F}).

Next we calculate $\eta_1$ in Eq.~(\ref{eq:eta_1}) as 
\begin{align}
\eta_1 &= \frac{n\zeta}{4} \frac{\int d\mathbf{r}\, r_y^2 e^{- \beta U(r,1)}}
{\int d\mathbf{r}\, e^{- \beta U(r,1)}} 
= \frac{n\zeta}{12} \frac{\int_0^\infty dr\, r^4 e^{- \beta U(r,1)}}
{\int_0^\infty dr\, r^2 e^{- \beta U(r,1)}}.
\end{align}
For a harmonic potential, the integration of $r^m$ can be generally expressed as
\begin{align}
g_m(p,q) &= \int_0^\infty dr\, r^m e^{-p(r-q)^2} \nonumber 
\\
& = \int_{-q}^\infty du\, (u+q)^m e^{-pu^2} \nonumber
\\
& =\sum_{n=0}^m \frac{m!}{(m-n)!n!}q^{m-n} \int_{-q}^\infty du\, u^n e^{-pu^2}.
\end{align}
The last integral can be further performed as follows.
\begin{align}
& \int_{-q}^0 du\, u^n e^{-pu^2}+
\int_{0}^\infty du\, u^n e^{-pu^2} 
\nonumber \\
& = \frac{p^{-(n+1)/2}}{2} \left[ 
(-1)^n \int_0^{pq^2} dt\, t^{(n+1)/2-1}e^{-t} \right.
\nonumber \\
&\left. + \int_0^\infty dt\, t^{(n+1)/2-1}e^{-t}
\right], \nonumber \\
& = \frac{p^{-(n+1)/2}}{2} \left[ 
[1+(-1)^n ] \int_0^{\infty} dt\, t^{(n+1)/2-1}e^{-t} \right.
\nonumber \\
& \left. -(-1)^n \int_{pq^2}^\infty dt\, t^{(n+1)/2-1}e^{-t} \right].
\end{align}
Finally, $g_m(p,q)$ becomes 
\begin{align}
& g_m(p,q) =\frac{1}{2}\sum_{n=0}^m \frac{m!}{(m-n)!n!} p^{-(n+1)/2}q^{m-n} \nonumber\\
&\times \left[ \left[ 1+(-1)^n \right] \Gamma\left(\frac{n+1}{2}\right)  -(-1)^n\Gamma
\left( \frac{n+1}{2}, pq^2 \right) \right], 
\label{eq:ap_g}
\end{align}
where $\Gamma(x)=\int_0^\infty dt\, t^{x-1}e^{-t}$ and 
$\Gamma(x,\alpha)=\int_\alpha^\infty dt\, t^{x-1}e^{-t}$
are the gamma function and the incomplete gamma function of the second kind, respectively~\cite{Abramowitz1972}.

\section{Hydrodynamic interactions between two spheres}
\label{app:hydro}

In this Appendix, we present the calculation of Eq.~(\ref{eq:h}) for the Fraenkel dimer model.
With the assumption that the fluid is isotropic, the Oseen tensor becomes $\delta_{\alpha\beta}/(6\pi\eta_{\rm s}r)$.
Substituting it into Eq.~(\ref{eq:h}) yields
\begin{align}
h=\frac{1}{6\pi\eta_{\rm s}}
\frac{\int d\textbf{r}\, \psi_0(r) / r}{\int d\textbf{r}\, \psi_0(r)}.
\end{align}
By taking $m=1,2$ in $g_m(p,q)$, Eq.~(\ref{eq:ap_g}), the dimensionless combination $\zeta h$ is obtained as 
\begin{align}
\zeta h(\epsilon)&= \frac{a}{\ell_0} \frac{g_1(\epsilon,1)}{g_2(\epsilon,1)}\nonumber\\
&=\frac{a}{\ell_0}\frac{ e^{-\epsilon} +
\sqrt{\pi\epsilon} \left[1+{\rm erf}\left(\sqrt{\epsilon }\right)\right]}
{e^{-\epsilon}+\sqrt{\pi\epsilon}[1+1/(2\epsilon) ] \left[ 1+ \rm{erf}\left(\sqrt{\epsilon }\right)\right]}.
\end{align}

For large dimers, $a/\ell_0\ll1$, the hydrodynamic effects become negligible.
The limiting behavior of $h$ for the Hookean, $\epsilon\ll1$, and stiff Fraenkel dimers, $\epsilon\gg1$, is given, respectively, by
\begin{align}
\zeta h(\epsilon) = \left\{ \begin{array}{lr}{\displaystyle \frac{2a}{\ell_0}\sqrt{\frac{\epsilon}{\pi}} } & {~~~\epsilon\ll1}, \\[2.0ex]
 {\displaystyle \frac{a}{\ell_0}\frac{1}{1+1/(2\epsilon)} } & {~~~\epsilon\gg1}.
\end{array} \right.
\label{eq:zetah}
\end{align}



\begin{thebibliography}{99}

\bibitem{Albertsbook}
B. Alberts, A. Johnson, P. Walter, J. Lewis, and M. Raff,
\textit{Molecular Biology of the Cell}
(Garland Science, New York, 2008).

\bibitem{Gerstein1994}
M. Gerstein, A. M. Lesk, and C. Chothia, 
Biochemistry \textbf{33}, 6739 (1994).

\bibitem{Togashi2007}
Y. Togashi and A. S. Mikhailov, 
Proc. Natl. Acad. Sci. (USA) \textbf{104}, 8697 (2007).

\bibitem{Sakaue2010}
T. Sakaue, R. Kapral, and A. S. Mikhailov, 
Eur. Phys. J. B \textbf{75}, 381 (2010).

\bibitem{Echeverria2011}
C. Echeverria, Y. Togashi, A. S. Mikhailov, and R. Kapral, 
Phys. Chem. Chem. Phys. \textbf{13}, 10527 (2011).

\bibitem{Aviram2018}
H. Y. Aviram, M. Pirchi, H. Mazal, Y. Barak, I. Riven, and G. Haran, 
Proc. Natl. Acad. Sci. (USA) \textbf{115},  3243 (2018).

\bibitem{Zhang2019}
Y. Zhang and H. Hess,
ACS Cent. Sci. \textbf{5}, 939 (2019).

\bibitem{Muddana2010}
H. S. Muddana, S. Sengupta, T. E. Mallouk, A. Sen, and P. J. Butler, 
J. Am. Chem. Soc. \textbf{132}, 2110 (2010).

\bibitem{Riedel2015}
C. Riedel, R. Gabizon, C. A. M. Wilson, K. Hamadani, K. Tsekouras, S. Marqusee, S. Press\'{e}, and C. Bustamante, 
Nature \textbf{517}, 227 (2015).

\bibitem{Illien2017_2}
P. Illien, X. Zhao, K. K. Dey, P. J. Butler, A. Sen, and R. Golestanian, 
Nano Lett. \textbf{17}, 4415 (2017).

\bibitem{Sengupta2013}
S. Sengupta, K. K. Dey, H. S. Muddana, T. Tabouillot, M. E. Ibele, P. J. Butler, and A. Sen, 
J. Am. Chem. Soc. \textbf{135}, 1406 (2013).

\bibitem{Jee2018}
A.-Y. Jee, S. Dutta, Y.-K. Cho, T. Tlusty, and S. Granick, 
Proc. Natl. Acad. Sci. (USA) \textbf{115}, 14 (2018).

\bibitem{Dey2016}
K. K. Dey, F. Y. Pong, J. Breffke, R. Pavlick, E. Hatzakis, C. Pacheco, and A. Sen, 
Angew. Chem. Int. Ed. \textbf{55}, 1113 (2016).

\bibitem{Zhao2017}
X. Zhao, K. K. Dey, S. Jeganathan, P. J. Butler, U. M. C\'{o}rdova-Figueroa, and A. Sen, 
Nano Lett. \textbf{17}, 4807 (2017).

\bibitem{Illien2017}
P. Illien, T. Adeleke-Larodo, and R. Golestanian, 
EPL \textbf{119}, 40002 (2017).

\bibitem{Adeleke-Larodo2019}
T. Adeleke-Larodo, P. Illien, and R. Golestanian, 
Eur. Phys. J. E \textbf{42}, 39 (2019).

\bibitem{Golestanian2015}
R. Golestanian, 
Phys. Rev. Lett. \textbf{115}, 108102 (2015).

\bibitem{Mikhailov2015}
A. S. Mikhailov and R. Kapral, 
Proc. Natl. Acad. Sci. (USA) \textbf{112}, E3639 (2015).

\bibitem{Kapral2016}
R. Kapral and A. S. Mikhailov,
Physica D \textbf{318}-\textbf{319}, 100 (2016).

\bibitem{Hosaka2017}
Y. Hosaka, K. Yasuda, R. Okamoto, and S. Komura, 
Phys. Rev. E \textbf{95}, 052407 (2017).

\bibitem{Xu2019}
M. Xu, J. L. Ross, L. Valdez, and A. Sen,
Phys. Rev. Lett. \textbf{123}, 128101 (2019).

\bibitem{Zhang2018}
Y. Zhang, M. J. Armstrong, N. M. B. Kazeruni, and H. Hess,
Nano Lett. \textbf{18}, 8025 (2018).
 
 \bibitem{Guenther2019}
J.-P. G\"{u}nther, G. Majer, and P. Fischer,
J. Chem. Phys. \textbf{150}, 124201 (2019).

\bibitem{Armoskaite2012}
V. Armo\v{s}kait\.{e}, K. Ramanauskien\.{e}, and V. Briedis,
Afr. J. Pham. Pharmacol. \textbf{6}, 1685 (2012).

\bibitem{Flechsig2019}
H. Flechsig and A. S. Mikhailov,
J. R. Soc. Interface \textbf{16}, 20190244 (2019).

\bibitem{Hosaka2019}
Y. Hosaka, S. Komura, and A. S. Mikhailov, unpublished.

\bibitem{Birdbook}
R. B. Bird, R. C. Armstrong, O. Hassager, and C. F. Curtiss, 
\textit{Dynamics of Polymeric Liquids, Vol. 2} (Wiley, New York, 1987).

\bibitem{Doibook}
M. Doi, \textit{Soft Matter Physics} (Oxford University, Oxford, 2013).

\bibitem{Fraenkel1952}
G. K. Fraenkel, J. Chem. Phys. \textbf{20}, 642 (1952).

\bibitem{Bird1997}
R. B. Bird, C. F. Curtiss, and K. J. Beers, 
Rheol. Acta. \textbf{36}, 269 (1997).

\bibitem{Abramowitz1972}
M. Abramowitz and I.A. Stegun, \textit{Handbook of Mathematical
Functions} (Dover, New York, 1972).

\bibitem{Kou2005}
S. C. Kou, B. J. Cherayil, W. Min, B. P. English, and X. S. Xie, 
J. Phys. Chem. \textbf{109}, 19068 (2005).

\bibitem{English2006}
B. P. English, W. Min, A. M. van Oijen, K. T. Lee, G. Luo, H. Sun, B. J. Cherayil, S. C. Kou, and X. S. Xie,
Nature Chem. Bio. \textbf{2}, 87 (2006).

\bibitem{Michaelis1913}
L. Michaelis and M. L. Menten,
Biochem. Z. \textbf{49}, 333-369 (1913).

\bibitem{Cao2011}
J. Cao, 
J. Phys. Chem. B \textbf{115}, 5493 (2011).

\bibitem{vanKampenbook}
N. G. van Kampen, 
\textit{Stochastic processes in physics and chemistry} (Elsevier Science, New York, 1992).

\bibitem{Lu1998}
H. P. Lu, L. Xun, and X. S. Xie, 
Science \textbf{282}, 1877 (1998).

\bibitem{Xie2001}
S. Xie, 
Single Mol. \textbf{2}, 229 (2001).

\bibitem{Saha2011}
S. Saha, S. Ghose, R. Adhikari, and A. Dua,
Phys. Rev. Lett. \textbf{107}, 218301 (2011).

\bibitem{Grima2009}
R. Grima,
Phys. Rev. Lett. \textbf{102}, 218103 (2009).

\bibitem{Oppenheimer2009}
N. Oppenheimer and H. Diamant, 
Biophys. J. \textbf{96}, 3041 (2009).

\bibitem{Oppenheimer2010}
N. Oppenheimer and H. Diamant, 
Phys. Rev. E \textbf{82}, 041912 (2010).

\bibitem{Bird1971}
R. B. Bird and H. R. Warner,
Trans. Soc. Rheol. \textbf{15}, 741 (1971).

\bibitem{Doibook2}
M. Doi and S. F. Edwards, 
\textit{The Theory of Polymer Dynamics}
(Oxford University, New York, 1986).

\bibitem{Warner1972}
H. R. Warner,
Ind. Eng. Chem. Fundam., \textbf{11}, 379 (1972).

\bibitem{Haenggi1990}
P. H\"{a}nggi, P. Talkner, and M. Brokovec, 
Rev. Mod. Phys. \textbf{62}, 251 (1990).


\end{thebibliography}
\end{document}